\documentstyle{mn} 

\input psfig 
 
\newcommand{\gta}{\mathrel{\hbox{\rlap{\lower.55ex \hbox {$\sim$}} 
                   \kern-.3em \raise.4ex \hbox{$>$}}}} 
\newcommand{\lta}{\mathrel{\hbox{\rlap{\lower.55ex \hbox {$\sim$}} 
                   \kern-.3em \raise.4ex \hbox{$<$}}}} 
 
\newif\ifAMStwofonts 
 
\ifoldfss 
  \ifCUPmtlplainloaded \else 
    \NewTextAlphabet{textbfit} {cmbxti10} {} 
    \NewTextAlphabet{textbfss} {cmssbx10} {} 
    \NewMathAlphabet{mathbfit} {cmbxti10} {} % for math mode 
    \NewMathAlphabet{mathbfss} {cmssbx10} {} %  "   "    " 
  \fi 
  \ifAMStwofonts 
    \ifCUPmtlplainloaded \else 
      \NewSymbolFont{upmath} {eurm10} 
      \NewSymbolFont{AMSa} {msam10} 
      \NewMathSymbol{\upi}     {0}{upmath}{19} 
      \NewMathSymbol{\umu}     {0}{upmath}{16} 
      \NewMathSymbol{\upartial}{0}{upmath}{40} 
      \NewMathSymbol{\leqslant}{3}{AMSa}{36} 
      \NewMathSymbol{\geqslant}{3}{AMSa}{3E}

       \let\le=\leqslant 
        
    \fi 
  \fi 
\fi % End of OFSS 
 
\ifnfssone 
  \newmathalphabet{\mathit} 
  \addtoversion{normal}{\mathit}{cmr}{m}{it} 
  \addtoversion{bold}{\mathit}{cmr}{bx}{it} 
  \newmathalphabet{\mathbfit} % math mode version of \textbfit{..} 
  \addtoversion{normal}{\mathbfit}{cmr}{bx}{it} 
  \addtoversion{bold}{\mathbfit}{cmr}{bx}{it} 
  \newmathalphabet{\mathbfss} % math mode version of \textbfss{..} 
  \addtoversion{normal}{\mathbfss}{cmss}{bx}{n} 
  \addtoversion{bold}{\mathbfss}{cmss}{bx}{n} 
  \ifAMStwofonts 
    \ifCUPmtlplainloaded \else 
      \UseAMStwoboldmath 
      \makeatletter 
      \new@mathgroup\upmath@group 
      \define@mathgroup\mv@normal\upmath@group{eur}{m}{n} 
      \define@mathgroup\mv@bold\upmath@group{eur}{b}{n} 
      \edef\UPM{\hexnumber\upmath@group} 
      \new@mathgroup\amsa@group 
      \define@mathgroup\mv@normal\amsa@group{msa}{m}{n} 
      \define@mathgroup\mv@bold\amsa@group{msa}{m}{n} 
      \edef\AMSa{\hexnumber\amsa@group} 
      \makeatother 
      \mathchardef\upi="0\UPM19 
      \mathchardef\umu="0\UPM16 
      \mathchardef\upartial="0\UPM40 
      \mathchardef\leqslant="3\AMSa36 
      \mathchardef\geqslant="3\AMSa3E 

       \let\le=\leqslant 

    \fi 
  \fi 
\fi % End of NFSS release 1 
\ifnfsstwo 
  \DeclareMathAlphabet{\mathbfit}{OT1}{cmr}{bx}{it} 
  \SetMathAlphabet\mathbfit{bold}{OT1}{cmr}{bx}{it} 
  \DeclareMathAlphabet{\mathbfss}{OT1}{cmss}{bx}{n} 
  \SetMathAlphabet\mathbfss{bold}{OT1}{cmss}{bx}{n} 
  \ifAMStwofonts 
    \ifCUPmtlplainloaded \else 
      \DeclareSymbolFont{UPM}{U}{eur}{m}{n} 
      \SetSymbolFont{UPM}{bold}{U}{eur}{b}{n} 
      \DeclareSymbolFont{AMSa}{U}{msa}{m}{n} 
      \DeclareMathSymbol{\upi}{0}{UPM}{"19} 
      \DeclareMathSymbol{\umu}{0}{UPM}{"16} 
      \DeclareMathSymbol{\upartial}{0}{UPM}{"40} 
      \DeclareMathSymbol{\leqslant}{3}{AMSa}{"36} 
      \DeclareMathSymbol{\geqslant}{3}{AMSa}{"3E} 

       \let\le=\leqslant 

    \fi 
  \fi 
\fi % End of NFSS release 2 
 
\ifCUPmtlplainloaded \else 
  \ifAMStwofonts \else % If no AMS fonts 
    \def\upi{\pi} 
    \def\umu{\mu} 
    \def\upartial{\partial} 
  \fi 
\fi 
\def\aaa#1{{A\&A,} {#1}}

\def\apj#1{{ApJ,} {#1}} 
 
\def\mnras#1{{MNRAS,} {#1}}

\newbox\grsign \setbox\grsign=\hbox{$>$} 
\newdimen\grdimen \grdimen=\ht\grsign 
\newbox\laxbox \newbox\gaxbox 
\setbox\gaxbox=\hbox{\raise.5ex\hbox{$>$}\llap 
     {\lower.5ex\hbox{$\sim$}}}\ht1=\grdimen\dp1=0pt 
\setbox\laxbox=\hbox{\raise.5ex\hbox{$<$}\llap 
     {\lower.5ex\hbox{$\sim$}}}\ht2=\grdimen\dp2=0pt 
\def\gax{\mathrel{\copy\gaxbox}} 
 
\def\lax{\mathrel{\copy\laxbox}} 
\def\lta{\lax} 
\def\gta{\gax} 
\title[Absence of winds in ADAFs] 
  {On the absence of winds in ADAFs} 
 
\author[Abramowicz, Lasota \& Igumenshchev] 
  {Marek A. Abramowicz$^{1,2,3}$\thanks{E-mail: 
   marek@fy.chalmers.se (MAA), lasota@iap.fr (JPL), ivi@fy.chalmers.se (IVI)}, 
  Jean-Pierre Lasota$^{1\star}$\thanks{On leave from:  
  Institut d'Astrophysique      
     de Paris, CNRS, 98bis Boulevard Arago, 75014 Paris, France}\thanks{also: 
     DARC, Observatoire de Paris, France}  
     \&\ Igor V. Igumenshchev$^{4,2\star}$\\ 
  $^1$ Institute for Theoretical Physics, University of California,  
  Santa Barbara, CA 93106-4030, USA\\ 
  $^2$ Institute for Theoretical Physics, G\"oteborg University and  
  Chalmers University     
     of Technology, 412 96 G\"oteborg, Sweden\\ 
  $^3$ Laboratorio Interdisciplanare SISSA and ICTP, Trieste, Italy\\ 
  $^4$ Institute of Astronomy, 48 Pyatnitskaya street, Moscow, 109017,  
  Russia\\ 
  } 
 
\pagerange{\pageref{firstpage}--\pageref{lastpage}} 
\pubyear{1998}

\begin{document} 
 
\maketitle 
 
\label{firstpage} 
 
\begin{abstract} 
 
We show that recently published assertions that advection dominated 
accretion flows (ADAFs) require the presence of strong winds are 
unfounded because they assume that low radiative efficiency in flows 
accreting at low rates onto black holes implies vanishing radial energy 
and angular momentum fluxes through the flow (which is also formulated 
in terms of the `Bernoulli function' being positive). This, however, 
is a property only of self-similar solutions which are an inadequate 
representation of global accretion flows. We recall general properties 
of accretion flows onto black holes and show that such, necessarily 
transonic, flows may have either positive or negative Bernoulli function 
depending on the flow viscosity.  
Flows with low viscosities  
($\alpha \lta 0.1$ in the $\alpha$--viscosity model)  
have a negative Bernoulli function. Without exception,  
all 2-D and 1-D numerical models of low viscosity flows constructed  
to date experience no significant outflows.  
At high viscosities the presence of outflows 
depends on the assumed viscosity, the equation of state and on the 
outer boundary condition. The positive sign of the Bernoulli function 
invoked in this context is irrelevant to the presence of 
outflows. As an illustration, we recall 2-D numerical models with 
moderate viscosity that have positive values of the Bernoulli function 
and experience no outflows. ADAFs, therefore, do not differ from this 
point of view from thin Keplerian discs: they may have, but they do not 
have to have strong winds. 
\end{abstract} 
 
 
\section{Introduction} 
 
In many systems containing accreting galactic and 
extragalactic black holes, the luminosity deduced from observations is 
much lower than the one obtained by assuming a `standard' radiation efficiency 
of $\sim 0.1$.  
It has been proposed that accretion in such systems can be modeled by  
advectively dominated accretion flows -- ADAFs (for recent reviews see  
%Kato, Inagaki, Mineshige \& Fukue 1996;  
Kato, Fukue \& Mineshige 1998; 
Abramowicz, Bj{\"o}rnsson \& Pringle 1998; Lasota 1999).  
In ADAFs, the main cooling process is advection of heat~--- 
radiative cooling is only a small perturbation in the energy balance 
and has no dynamical importance. ADAFs are quite successful in 
explaining observed spectral properties of accretion onto black hole in low 
mass X-ray binaries, Galactic center, and some active galactic nuclei. 
 
Recently, Blandford \& Begelman (1999, hereafter BB99) put in question 
the physical self-consistency of ADAF models by arguing that flows with 
small radiative efficiency should experience outflows, which 
in some cases could
be so strong as to prevent accretion of matter onto the black hole. 
Low luminosities, therefore, would not be due to low radiative efficiency, but 
simply to absence of accreting matter near the central object. 
Hence, ADAFs should be replaced by what BB99 call ADIOs, 
i.e. `advection-dominated inflows-outflows' (later 
`advective' has been changed to `adiabatic'). The BB99 argument about the 
necessity of outflows from ADAFs is based on two statements  
about what they call the `Bernoulli constant' (the significance of this 
`constant' - in reality a function - is discussed in 
the next Section): [1]~The Bernoulli constant in flows with low 
radiative efficiency must always be positive, a point which was first 
made by Narayan \& Yi (1994), [2]~A positive Bernoulli constant implies 
outflows. 
 
In the present paper we show that [1]~the vertically integrated 
Bernoulli function is {\it everywhere 
negative} in small viscosity ADAFs which have a vanishing viscous torque  
at the flow inner `edge',   
and are matched to the standard thin Shakura-Sunyaev disc (SSD) 
at the outer edge,  
[2]~A positive Bernoulli function in an ADAF does not imply  
outflows~--- in a representative class of 2-D numerical models of ADAFs 
with moderate viscosity, Bernoulli function may be positive but despite of that 
outflows are always absent.  
 
The main conclusion of our paper is that while it is not yet 
clear whether some ADAFs with high viscosity could indeed have 
significant outflows, both general theoretical arguments and numerical 
simulations point out that it is unlikely that ADAFs with low viscosity 
could experience even moderate outflows of {\sl purely hydrodynamical} 
origin. 
Similar conclusion has been recently obtained by Nakamura (1998), 
who used a different approach. 
 
We use here cylindrical coordinates $(r, z, \phi)$ and 
denote gravitational radius of the accreting black hole (with the mass 
$M$) by $r_G = 2GM/c^2$. We model the gravitational field of the black 
hole by the Paczy{\'n}ski \& Wiita (1980) potential.  
We assume in this paper that viscosity is small. 
%In terms of the `standard $\alpha$ prescription' 
This means  that $\alpha\la 0.1$ for the kinematic 
viscosity coefficient given by a phenomenological formula
%%%%%%%%%%%%%%%%%%%%%%%%%%%%%%%%%%%%%%%%%%%%%%%%%%%%%%%%%%%%%%%%%%%%%%%%%% 
$$ \nu = \alpha {\ell}_P C_{sound}, \eqno (1.1) $$ 
%%%%%%%%%%%%%%%%%%%%%%%%%%%%%%%%%%%%%%%%%%%%%%%%%%%%%%%%%%%%%%%%%%%%%%%%%% 
where ${\ell}_P = P/|\nabla P|$ is the pressure length scale and $C_{sound}$ is 
the sound speed. 
Our theoretical arguments do not depend on a particular 
functional form (prescription) of $\alpha$. 
In numerical models we assume the `standard' prescription, $\alpha=const$.
Recent theoretical estimates based on numerical simulations of turbulent 
viscosity (see e.g. Balbus, Hawley \& Stone 1996) show that  
most likely 
$\alpha \la 0.1$. In this paper we shall consider mainly this {\it 
low viscosity} range of $\alpha$. Arguments based on fitting predicted 
spectral properties of ADAF models to observations (see Narayan 1999 
for review) seem to require the {\it moderate viscosity} range, $0.1 \la 
\alpha \la 0.3$, that partially overlaps with the low viscosity range 
considered in this paper. As pointed out by Rees (private communication) it is not clear if the {\it high viscosity} range, $\alpha \ga 0.3$, is physically realistic.

\section{The Bernoulli constant, function and parameter} 
 
In stationary, inviscid flows with no energy sources or losses, the 
quantity, 
%%%%%%%%%%%%%%%%%%%%%%%%%%%%%%%%%%%%%%%%%%%%%%%%%%%%%%%%%%%%%%%%%%%%%%%%%% 
$$ B_0 = W + {1\over 2}V^2 + \Phi = const,  \eqno (2.1)$$ 
%%%%%%%%%%%%%%%%%%%%%%%%%%%%%%%%%%%%%%%%%%%%%%%%%%%%%%%%%%%%%%%%%%%%%%%%%% 
is constant along each individual stream line, but, in general, is different for 
different stream lines. This quantity is called the  Bernoulli constant. Here $W$ is the specific enthalpy, $V$ is the velocity (all three components 
included), and $\Phi$ is the gravitational potential. Obviously, a 
particular streamline may end up at infinity {\it only if} $B_0 > 0$ 
along it. The existence of stream lines with $B_0 > 0$ is therefore a 
{\it necessary} condition for outflows in stationary inviscid flows 
with no energy sources or losses, and $B_0<0$ for all streamlines is a 
{\it sufficient} condition for the absence of outflows. However, $B_0>0$ 
is {\it not} a sufficient condition for outflows. For example, in the 
case of the classical Bondi's accretion (pure inflow) $B_0$ 
is a universal positive constant. 
 
In all viscous flows, $B_0$ defined by (2.1) is, of 
course, {\it not} constant along individual stream lines. However, 
the so-called `Bernoulli parameter', 
%%%%%%%%%%%%%%%%%%%%%%%%%%%%%%%%%%%%%%%%%%%%%%%%%%%%%%%%%%%%%%%%%%%%%%%%%%%% 
$$ {\tilde B}_0 =  
{1 \over V_K^2} \left [ W + {1\over 2}V^2 + \Phi \right ],  
\eqno (2.2)$$ 
%%%%%%%%%%%%%%%%%%%%%%%%%%%%%%%%%%%%%%%%%%%%%%%%%%%%%%%%%%%%%%%%%%%%%%%%%%%%  
is a universal constant in 1-D, vertically integrated, self-similar 
Newtonian models of ADAFs introduced by Narayan \& Yi (1994).  
Here $V_K$ is the Keplerian velocity. The reason for that is simple: in  
self-similar models all quantities scale as some power of the 
cylindrical coordinate $r$. In particular, because both  
%%%%%%%%%%%%%%%%%%%%%%%%%%%%%%%%%%%%%%%%%%%%%%%%%%%%%%%%%%%%%%%%%%%%%%%%%%%%% 
$$ B = W + {1\over 2}V^2 + \Phi \not= const, \eqno (2.3)$$  
%%%%%%%%%%%%%%%%%%%%%%%%%%%%%%%%%%%%%%%%%%%%%%%%%%%%%%%%%%%%%%%%%%%%%%%%%%%%% 
and $V_K^2$ scale as $r^{-1}$, their ratio ${\tilde B}_0$ must be a constant. 
Narayan \& Yi (1994) and BB99 have argued that because the 
vertically integrated models have ${\tilde B}_0 >0$, in generic 2-D flows 
there should be streamlines escaping to infinity. 
 
This argument is not correct for the simple reason that neither $B_0 > 0$ (as 
already mentioned), nor ${\tilde B}_0 > 0$, are sufficient conditions for 
outflow existence. Note that ${\tilde B}_0= {\rm constant}$ is not 
a real physical conservation law, but rather an artifact induced by a
purely mathematical assumption that the flow is self-similar. This {\it 
very strong} assumption is motivated only by practical convenience, 
not by the physical properties of a flow.  Indeed, in all numerical, global 
models of ADAFs constructed so far, self-similarity does not represent 
well the global flow properties 
and ${\tilde B}_0$ changes its value and even its sign 
from place to place. Large scale changes are due to the 
global balance between viscous heating, advective cooling and $PdV$ 
work. Small scale changes are typical for ADAFs in which a 
strong convection (circulation) is present  
(these ADAFs have a moderate or small 
viscosity). In this case, the sign of ${\tilde B}_0$ changes rather 
abruptly, tracing convective bubbles. In addition, at a given place, 
the value and the sign of ${\tilde B}_0$ strongly fluctuates in time. 
 
The Bernoulli constant $B_0$ is not a useful quantity for ADAF study,  
because they are viscous flows. The Bernoulli parameter ${\tilde B}_0$ has no 
physical significance and therefore it is not a convenient quantity for 
theoretical arguments. We shall use here the `Bernoulli 
function', which is formally defined by (2.3), but~--- of course~--- 
in ADAFs (or any other viscous flows) it is {\it not} constant along 
streamlines.  
As in the case of steady dissipation-free flows, $B < 0$ 
everywhere is a sufficient condition for the absence of outflows. 
 
We were forced to introduce such new object 217 years after 
the death of Daniel Bernoulli because of the above-mentioned articles on outflows from ADAFs, which use the related concepts of the Bernoulli constant 
or `parameter'. It is easier, as shown below, to refute their arguments by using the same `paradigm'.

\section{The inner boundary condition} 
 
In this Section we would like to stress the importance of the fact that the 
accreting body is a black hole, in particular, the implications of the transonic nature of the accretion flow. 
 
We will assume that the flow viscosity is low. Paczy{\'n}ski (1978, 
unpublished) noticed that in this case the mass loss from 
the inner part of an accretion disc around a black hole is fully governed by the general relativistic 
effect of `relativistic Roche lobe overflow' which operates close to 
a  `cusp' radius $r_{cusp}$, where the angular momentum of 
the disc takes the Keplerian value, $j(r_{cusp}) = j_K(r_{cusp})$ (see 
Abramowicz 1981, 1985 for details). Here $j_K=V_K r$ is the  
Keplerian angular momentum. The cusp is formed by a critical equipotential 
surface, similar (but topologically different) to the critical Roche surface in the binary stellar-system problem. 
 
Jaroszy{\'n}ski, Abramowicz \& Paczy{\'n}ski (1980) proved that for 
small viscosity accretion discs  one must have $r_{mb} < r_{cusp} < r_{ms}$, 
where $r_{mb}$ and $r_{ms}$ are respectively the radii of the marginally bound, and  the marginally stable Keplerian circular orbits around a 
black hole. In the case of a non-rotating (Schwarzschild) black hole, 
as well as in the Paczy\'nski \& Wiita (1980) potential, 
$r_{mb}=2r_G$ and $r_{ms}=3r_G$. At $r_{ms}$ the Keplerian angular 
momentum has a minimum value $j_{ms} = (3/2)^{3/2} r_G c$.  Numerical 
simulations show that low viscosity accretion discs with reasonable {\it 
outer} boundary conditions (discussed in the next Section) have always a 
{\it high specific angular momentum}, that is for $r_{cusp} < r < r_0$,  
they are always super-Keplerian:  
$j(r) > j_{K}(r)$, where $r_0 \approx 5 r_G$ (see Figure~1). 
 
The mass loss through the cusp, at the rate ${\dot M}_{cusp}$, induces an 
advective cooling ${Q}^-_{cusp}$ which is a very sensitive function of 
the vertical thickness $h$ of the disc at the cusp: independent of the 
equation of state (i.e. independent of the adiabatic index of the 
accreted matter $\gamma$) one has  
${Q}^-_{cusp} \sim \Sigma h^3$, 
where $\Sigma$ is the disc surface density at the cusp. This 
implies that the relativistic Roche lobe overflow stabilizes possible 
unstable thermal modes in the region close to $r_{cusp}$ because an 
overheating would cause vertical expansion, which then would induce strong advective 
cooling. More precisely, in terms of Pringle's thermal stability 
criterion (Pringle 1976; Piran 1978), 
%%%%%%%%%%%%%%%%%%%%%%%%%%%%%%%%%%%%%%%%%%%%%%%%%%%%%%%%%%%%%%%%%%%%%%%%%%%% 
$$ \left ( {{\partial \ln Q^+_{vis}} \over {\partial \ln h}}\right )_\Sigma < 
\left ( {{\partial \ln Q^-_{cusp}} \over {\partial \ln h}}\right )_\Sigma , 
\eqno (3.1)$$ 
%%%%%%%%%%%%%%%%%%%%%%%%%%%%%%%%%%%%%%%%%%%%%%%%%%%%%%%%%%%%%%%%%%%%%%%%%%%% 
where $Q^+_{vis} \sim \Sigma h^2$ is the rate of viscous heating, one has 
$2=(l.h.s) < (r.h.s)=3$, which proves thermal stability (Abramowicz, 
1981). Thus, an increase in the Bernoulli function caused by overheating 
does not produce outflows, but enhances the inflow into the black hole. 
This analytic prediction has been confirmed in all details by 2-D time 
dependent numerical simulations. In particular, Igumenshchev, Chen \& 
Abramowicz (1996) showed that the analytic formula (Abramowicz, 1985) for 
the rate of the mass inflow into the black hole induced by the 
relativistic Roche lobe overflow reproduces, in a wide range of 
parameters and with impressive {\it quantitative} accuracy, the behavior of 
${\dot M}_{cusp}$ calculated in {\it all} their numerical simulations. 
 
For low viscosity (high angular momentum) black hole accretion flows, 
Abramowicz \& Zurek (1981) found that the regularity condition at  
the sonic point $V_r^2 = C_{sound}^2$ requires 
%%%%%%%%%%%%%%%%%%%%%%%%%%%%%%%%%%%%%%%%%%%%%%%%%%%%%%%%%%%%%%%%%%%%%%%%%%%% 
$$ \left({C_{sound}\over c}\right)^2 \equiv \varepsilon^2 =  
{j_K^2(r_{sonic}) - j^2(r_{sonic}) \over 2 r_{sonic}^2 c^2}. 
 \eqno (3.2)$$ 
%%%%%%%%%%%%%%%%%%%%%%%%%%%%%%%%%%%%%%%%%%%%%%%%%%%%%%%%%%%%%%%%%%%%%%%%%%%% 
Because $\varepsilon \ll 1$ and $j_K(r_{sonic}) \approx 
r_{sonic}c$, one has 
%%%%%%%%%%%%%%%%%%%%%%%%%%%%%%%%%%%%%%%%%%%%%%%%%%%%%%%%%%%%%%%%%%%%%%%%%%%% 
$$ {j(r_{sonic}) \over j_K(r_{sonic})} 
= 1 - {r_{sonic}^2 c^2\over j_K^2(r_{sonic})}\varepsilon^2 
+{\cal O}(\varepsilon^4), 
\eqno (3.3)$$ 
%%%%%%%%%%%%%%%%%%%%%%%%%%%%%%%%%%%%%%%%%%%%%%%%%%%%%%%%%%%%%%%%%%%%%%%%%%%% 
and 
%%%%%%%%%%%%%%%%%%%%%%%%%%%%%%%%%%%%%%%%%%%%%%%%%%%%%%%%%%%%%%%%%%%%%%%%%%%% 
$$ {r_{sonic}\over r_{cusp}} = 1 -  
{\cal O}(\varepsilon^2), \eqno (3.4) $$ 
%%%%%%%%%%%%%%%%%%%%%%%%%%%%%%%%%%%%%%%%%%%%%%%%%%%%%%%%%%%%%%%%%%%%%%%%%%%% 
i.e. the sonic point is very close to the cusp. 
These properties have been confirmed by 1-D and 2-D numerical 
simulations performed independently by numerous authors. 
 
The supersonic flow at $r<r_{sonic}$ does not hydrodynamically 
influence the subsonic flow at $r>r_{sonic}$. For this reason, 
somewhere at an `inner edge' $r_{in} \approx r_{sonic} \approx 
r_{cusp}$, the viscous torque vanishes, $g(r_{in}) =0$.   
From (2.3) and (3.2)~-- (3.4) one derives, for a polytropic 
fluid, i.e. with $W = C^2_{sound}/(\gamma - 1)$, 
%%%%%%%%%%%%%%%%%%%%%%%%%%%%%%%%%%%%%%%%%%%%%%%%%%%%%%%%%%%%%%%%%%%%%%%%%%%% 
$$ {B(x_{sonic}) \over c^2} = -  
{{x_{sonic} - 2}\over {4(x_{sonic} - 1)^2}} + 
{{3 - \gamma} \over 2(\gamma - 1)}\varepsilon^2 +{\cal O}(\varepsilon^4). 
\eqno (3.5)$$ 
%%%%%%%%%%%%%%%%%%%%%%%%%%%%%%%%%%%%%%%%%%%%%%%%%%%%%%%%%%%%%%%%%%%%%%%%%%%% 
Here $x \equiv r/r_G$. In the physically relevant region, $2 \le 
x_{sonic} \le 3$, the leading term  
(zeroth order in $\varepsilon^2$) of 
this function equals, independently of $\gamma$,  
to {\it negative} binding energy at circular Keplerian orbit and varies 
between $0$ and $-1/16$.    
 
Thus, for ADAFs with low viscosity (high angular momentum) one {\it 
should} adopt the following inner boundary condition, 
%%%%%%%%%%%%%%%%%%%%%%%%%%%%%%%%%%%%%%%%%%%%%%%%%%%%%%%%%%%%%%%%%%%%%%%%%%% 
$$ g(r_{in}) =0,~~~~j(r_{in}) = j_K(r_{in}). \eqno (3.6)$$ 
%%%%%%%%%%%%%%%%%%%%%%%%%%%%%%%%%%%%%%%%%%%%%%%%%%%%%%%%%%%%%%%%%%%%%%%%%%% 
because this is imposed by fundamental properties of the black hole 
gravitational field and thus {\it must be} always obeyed. In addition, 
if at the sonic point $C_{sound}/c \ll 1$, the sonic point regularity 
condition demands that the Bernoulli function at the inner edge should be 
negative, 
%%%%%%%%%%%%%%%%%%%%%%%%%%%%%%%%%%%%%%%%%%%%%%%%%%%%%%%%%%%%%%%%%%%%%%%%%%% 
$$ B(r_{in}) \approx B(r_{sonic}) < 0, \eqno (3.7)$$ 
%%%%%%%%%%%%%%%%%%%%%%%%%%%%%%%%%%%%%%%%%%%%%%%%%%%%%%%%%%%%%%%%%%%%%%%%%%% 
independent of the equation of state (independent of $\gamma$).

\section{The outer boundary condition} 
 
ADAFs cannot exist for arbitrary large radii. For example,  
according to Abramowicz et al. (1995) ADAFs cannot extent beyond the
radius
%%%%%%%%%%%%%%%%%%%%%%%%%%%%%%%%%%%%%%%%%%%%%%%%%%%%%%%%%%%%%%%%%%%%%%%%%%% 
$$ r_{max} = C{\alpha^4 \over {\dot m}^2}r_G \gg r_{in}, \eqno (4.1)$$ 
%%%%%%%%%%%%%%%%%%%%%%%%%%%%%%%%%%%%%%%%%%%%%%%%%%%%%%%%%%%%%%%%%%%%%%%%%%% 
where ${\dot m} = {\dot M}/{\dot M}_{Edd}$ and $C \simeq 10^2$. 
In general, $C$ and the power of $\alpha$ depend on the cooling mechanisms 
included into the model (see Menou, Narayan \& Lasota 1999). 
 
Observations suggest that, in several systems the inner ADAF is surrounded 
by a geometrically thin, standard Shakura-Sunyaev disc so that there must exist 
a transition region where, for $r\sim r_{out}$  
the ADAF is matched to a Keplerian disc.  
The physical mechanism which triggers the transition is not known  
(see, however, Meyer \& Meyer-Hofmeister 1994; Kato \& Nakamura 1998)  
and it is not clear what is the relation between $r_{max}$ and $r_{out}$. 
 
These uncertainties make the conditions at the outer edge $r_{out}$  
of low viscosity accretion discs less precisely determined than the 
conditions at the inner edge. Despite of this, Abramowicz, Igumenshchev 
\& Lasota (1998) found a simple analytic argument that proves that 
independent of the physical reason for the transition, the angular 
momentum close at $r_{out}$ should have exactly the Keplerian value, 
$j(r_{out}) = j_K(r_{out})$. This was confirmed by numerical models of 
the only specific ADAF-SSD transition model worked out to date, in 
which the transition occurs due to the presence of a turbulent flux 
(Honma 1996, Kato \& Nakamura 1998). 
 
With an ADAF joining the SSD at the outer edge, one has, 
%%%%%%%%%%%%%%%%%%%%%%%%%%%%%%%%%%%%%%%%%%%%%%%%%%%%%%%%%%%%%%%%%%%%%%%%%% 
$$j(r_{out}) = j_K(r_{out}),~~~~B(r_{out}) \approx e_K(r_{out}) <0,  
\eqno (4.2)$$ 
%%%%%%%%%%%%%%%%%%%%%%%%%%%%%%%%%%%%%%%%%%%%%%%%%%%%%%%%%%%%%%%%%%%%%%%%%% 
Here $e_K(r_{out})$ is the {\it negative} Keplerian orbital binding 
energy. The second equation in (4.2) follows from 
%%%%%%%%%%%%%%%%%%%%%%%%%%%%%%%%%%%%%%%%%%%%%%%%%%%%%%%%%%%%%%%%%%%%%%%%%% 
$$B(r_{out}) = e_K(r_{out})\left[ 1 + 
{\cal O}\left( (h_{out}/r_{out})^2\right) \right] < 0,  \eqno (4.3)$$ 
%%%%%%%%%%%%%%%%%%%%%%%%%%%%%%%%%%%%%%%%%%%%%%%%%%%%%%%%%%%%%%%%%%%%%%%%%% 
where $h_{out} \ll r_{out}$ is the thickness of the outer SSD. 
In all numerical simulations of low viscosity, high 
angular momentum accretion flows, the outer boundary condition (4.2) 
is {\it approximately} fulfilled at all radii $r \gg r_{in}$ 
independent of whether the ADAF-SSD transition occurs.

\section{The angular momentum} 
 
%%%%%%%%%%%%%%%%%%%%%%%%%%%%%%%%%%%%%%%%%%%%%%%%%%%%%%%%%%%%%%%%%%%%%%%%%%%% 
%%%%%%%%%%%%%%%%%%%%%%%%%%%%%%%%%%%%%%%%%%%%%%%%%%%%%%%%%%%%%%%%%%%%%%%%%%%% 
\begin{figure} 
%\vspace{8.2cm} 
\hbox{\psfig{file=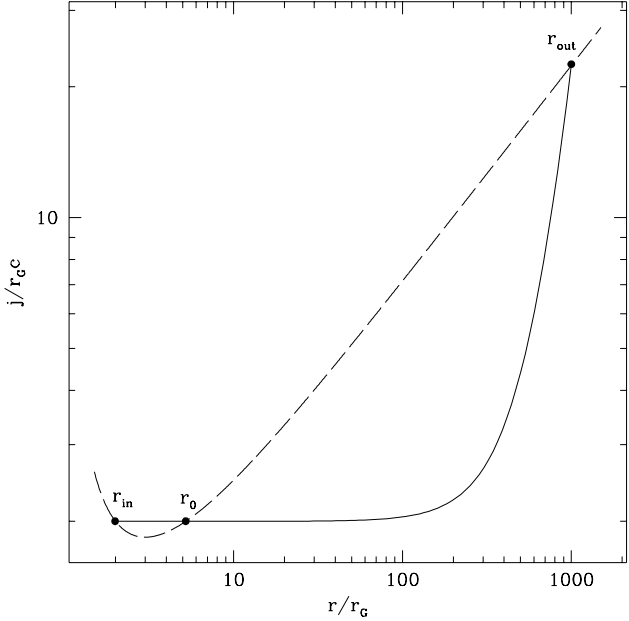,height=8.2cm}} 
\caption{ 
Angular momentum distribution for low viscosity ADAF (solid line) 
that fulfills inner (3.6) and outer (4.2) boundary conditions, 
given by Paczy{\'n}ski's fitting formula (5.2). The dashed 
line corresponds to the Keplerian distribution $j_K(r)$. 
} 
 \label{fig1} 
\end{figure} 
%%%%%%%%%%%%%%%%%%%%%%%%%%%%%%%%%%%%%%%%%%%%%%%%%%%%%%%%%%%%%%%%%%%%%%%%%%%% 
%%%%%%%%%%%%%%%%%%%%%%%%%%%%%%%%%%%%%%%%%%%%%%%%%%%%%%%%%%%%%%%%%%%%%%%%%%%% 
 
The shape of angular momentum distribution between $r_{in}$ and $r_{out}$ 
depends mainly on viscosity but, as numerous 1-D and 2-D models 
demonstrate, in the low viscosity case it is always similar to that shown 
in Figure~1.  
The physical reason for such a shape is clear. Close 
to the inner edge, the viscous torque is ineffective (note that 
$g(r_{in}) = 0$) and thus the specific angular momentum 
has a very small gradient. 
Thus, the location of $r_{in}$ roughly determines the location of the first 
crossing point $r_0$ between $j(r)$ and $j_K(r)$ curves, and therefore 
also the $1/r^3$ weighted area $A_{in}$ 
between these curves in the region $[r_{in}, r_0]$. 
Mechanical equilibrium condition demands that (Abramowicz, Calvani \& 
Nobili 1980), 
%%%%%%%%%%%%%%%%%%%%%%%%%%%%%%%%%%%%%%%%%%%%%%%%%%%%%%%%%%%%%%%%%%%%%%%%%%% 
$$ A_{in}=A_{out},   \eqno (5.1) $$ 
$$ A_{in}\equiv\int_{r_{in}}^{r_0} {{j^2(r) - j^2_K(r)} \over r^3}dr $$ 
$$ A_{out}\equiv\int_{r_0}^{r_{out}} {{j_K^2(r) - j^2(r)} \over r^3}dr, $$ 
%%%%%%%%%%%%%%%%%%%%%%%%%%%%%%%%%%%%%%%%%%%%%%%%%%%%%%%%%%%%%%%%%%%%%%%%%%% 
and thus the $1/r^3$ weighted area $A_{out}$ 
between these curves in the region  
$[r_0, r_{out}]$  
is also roughly determined. In addition, one should have $dj/dr > 
0$ and $d(j/r^2)/dr <0$. 
 
The function $j(r)$ can be approximated by the analytic fitting formula 
used in the Paczy{\'n}ski's (1998) toy model for ADAFs, 
%%%%%%%%%%%%%%%%%%%%%%%%%%%%%%%%%%%%%%%%%%%%%%%%%%%%%%%%%%%%%%%%%%%%%%%%%%% 
$$ j(r) = j_{r_{in}} \left[ 1 + b \left( { r \over r_{in}} - 1 \right) ^a 
+ b \left( { r \over r_{in} } - 1 \right) ^{3a} \right] ^{1.5/a}. 
\eqno (5.2)$$ 
%%%%%%%%%%%%%%%%%%%%%%%%%%%%%%%%%%%%%%%%%%%%%%%%%%%%%%%%%%%%%%%%%%%%%%%%%%% 
Note that this formula is different from the one 
in the published version of Paczy\' nski's article. 
The error it contained has been corrected in the astro-ph version. 
Here $a$ and $b$ are constants that depend on $r_{in}$ and $r_{out}$, and 
can be determined from (5.1). Although in actual models the range of radii 
for which the angular momentum is approximately constant is much reduced, 
one should stress, that (5.2) is a good qualitative 
representation of the {\sl generic} angular momentum distribution in ADAFs when the magnitude of viscosity is low {\it independent} of the functional form of viscosity.

\section{The Bernoulli function} 
 
Imagine a cylinder $r={\rm const}$ crossing an ADAF from its upper, 
$z=h(r)$, to lower $z=-h(r)$, surface. Let $\dot{M}_0$, $\dot{J}_0$ and 
$\dot{E}_0$ denote, respectively, the total amount of mass
($\dot{M}_0<0$), angular 
momentum and energy that cross the surface of the cylinder per unit 
time. In a steady state, from the Navier-Stockes equations of mass, 
angular momentum and energy conservations integrated along the 
cylinder it follows that, 
%%%%%%%%%%%%%%%%%%%%%%%%%%%%%%%%%%%%%%%%%%%%%%%%%%%%%%%%%%%%%%%%%%%%%%%%%% 
$${\dot M}_0 = \int_{-h(r)}^{+h(r)} 2\pi r \rho (r,z) V_r(r,z) dz = 
  {\rm const}, \eqno (6.1a)$$ 
%%%%%%%%%%%%%%%%%%%%%%%%%%%%%%%%%%%%%%%%%%%%%%%%%%%%%%%%%%%%%%%%%%%%%%%%%% 
$${\dot J}_0 = {\dot M}_0 j(r) + g(r) = {\rm const}, \eqno (6.1b)$$ 
%%%%%%%%%%%%%%%%%%%%%%%%%%%%%%%%%%%%%%%%%%%%%%%%%%%%%%%%%%%%%%%%%%%%%%%%%% 
$${\dot E}_0 = {\dot M}_0 B(r) + \Omega(r)g(r) = {\rm const}.  \eqno (6.1c)$$ 
%%%%%%%%%%%%%%%%%%%%%%%%%%%%%%%%%%%%%%%%%%%%%%%%%%%%%%%%%%%%%%%%%%%%%%%%%% 
Here $\Omega (r)=j(r)/r^2$ is the angular velocity. The radiative energy 
flux from ADAFs is very small and it was ignored in (6.1c).   
Derivation of (6.1) 
assumes that the flow has an azimuthal symmetry (no dependence on $\phi$), 
and that the orbital velocity $V_{\phi}$ is much greater than the 
accretion velocity $V_r$ (and `vertical' velocities) which is true, 
except very close to the inner 
edge, for flows with a small viscosity considered in this Section. 
Except $h(r)$, each radial function that appear in (6.1) represents the 
averaged value of the corresponding quantity. In particular, 
%%%%%%%%%%%%%%%%%%%%%%%%%%%%%%%%%%%%%%%%%%%%%%%%%%%%%%%%%%%%%%%%%%%%%%%%%% 
$$ B(r) = {1\over {2h(r){\dot M_0}}} 
\int_{-h(r)}^{+h(r)} 2\pi r \rho (r,z) V_r(r,z) B(r,z) dz. \eqno (6.2)$$ 
%%%%%%%%%%%%%%%%%%%%%%%%%%%%%%%%%%%%%%%%%%%%%%%%%%%%%%%%%%%%%%%%%%%%%%%%%% 
The same equations (6.1), with the same assumptions and with the same 
averaging procedure (6.2), have been used by BB99, and by Paczy{\'n}ski 
(1998) in his recent illuminating paper on a toy model of ADAFs. 
 
According to Narayan and Yi (1994) in self-similar models of ADAFs, both 
$j(r)$ and $g(r)$ scale as $r^{1/2}$, and both $B(r)$ and $\Omega(r)g(r)$ 
scale as $r^{-1}$. These scaling properties imply that ${\dot J} = 
C_Jr^{1/2}$ and ${\dot E} = C_Er^{-1}$, with $C_J={\rm const}$,  
$C_E={\rm const}$.   
Thus, the fluxes $\dot{J}_0$ and $\dot{E}_0$ can be constant if and only 
if $C_J = C_E = 0$. This implies that both the angular momentum flux, and 
the energy flux are exactly zero in self-similar models: $\dot{J}_0 = 0$, 
and $\dot{E}_0=0$ (BB99). From the first of these equations it follows 
that $g(r) = -j(r)/{\dot M}_0$. This, together with the second equation show that the Bernoulli function must be positive, 
%%%%%%%%%%%%%%%%%%%%%%%%%%%%%%%%%%%%%%%%%%%%%%%%%%%%%%%%%%%%%%%%%%%%%%%%%% 
$$B(r) = \Omega(r)j(r) = [r\Omega(r)]^2 > 0.  \eqno (6.3)$$ 
%%%%%%%%%%%%%%%%%%%%%%%%%%%%%%%%%%%%%%%%%%%%%%%%%%%%%%%%%%%%%%%%%%%%%%%%%% 
 
The conclusion $B(r) > 0$ is based on self-similar solutions which are 
obviously inconsistent with boundary conditions. We shall 
show that by taking properly into account the boundary conditions 
(3.6) and (4.2), one arrives at the opposite conclusion: $B(r) < 0$. 
  
The inner boundary condition $g(r_{in}) = 0$ implies that 
%%%%%%%%%%%%%%%%%%%%%%%%%%%%%%%%%%%%%%%%%%%%%%%%%%%%%%%%%%%%%%%%%%%%%%%%%% 
$${\dot M}_0 j(r_{in}) = {\dot M}_0 j(r) + g(r), \eqno (6.4a)$$ 
%%%%%%%%%%%%%%%%%%%%%%%%%%%%%%%%%%%%%%%%%%%%%%%%%%%%%%%%%%%%%%%%%%%%%%%%%% 
$${\dot M}_0 B(r_{in}) = {\dot M}_0 B(r) + \Omega(r)g(r).  \eqno 
(6.4b)$$ 
%%%%%%%%%%%%%%%%%%%%%%%%%%%%%%%%%%%%%%%%%%%%%%%%%%%%%%%%%%%%%%%%%%%%%%%%%% 
From the last two equations one derives 
%%%%%%%%%%%%%%%%%%%%%%%%%%%%%%%%%%%%%%%%%%%%%%%%%%%%%%%%%%%%%%%%%%%%%%%%%% 
$$ B(r) = B(r_{in}) + \Omega(r)j(r) - \Omega(r)j(r_{in}).  \eqno 
(6.5)$$ 
%%%%%%%%%%%%%%%%%%%%%%%%%%%%%%%%%%%%%%%%%%%%%%%%%%%%%%%%%%%%%%%%%%%%%%%%%% 
Equation (6.5) yields, at the outer edge of the disc $r_{out}$  
%%%%%%%%%%%%%%%%%%%%%%%%%%%%%%%%%%%%%%%%%%%%%%%%%%%%%%%%%%%%%%%%%%%%%%%%%% 
$$ B(r_{in}) = B(r_{out}) - \Omega(r_{out})[ j(r_{out}) - 
j(r_{in})]. \eqno (6.6)$$ 
%%%%%%%%%%%%%%%%%%%%%%%%%%%%%%%%%%%%%%%%%%%%%%%%%%%%%%%%%%%%%%%%%%%%%%%%%% 
Because in stable discs $j(r_{out}) > j(r_{in})$, the last term on the 
right hand side of this equation is always negative.  
From this fact and from equation (4.3) one concludes that for standard 
ADAFs, i.e. for those that have a vanishing torque at the inner edge 
$r_{in}$ and match the standard Shakura-Sunyaev disc at the outer edge 
$r_{out}$ one must have 
%%%%%%%%%%%%%%%%%%%%%%%%%%%%%%%%%%%%%%%%%%%%%%%%%%%%%%%%%%%%%%%%%%%%%%%%%% 
$$ B(r_{in}) < B(r_{out}) < 0.  \eqno (6.7) $$ 
%%%%%%%%%%%%%%%%%%%%%%%%%%%%%%%%%%%%%%%%%%%%%%%%%%%%%%%%%%%%%%%%%%%%%%%%%% 
Thus, the Bernoulli function in standard ADAFs must be negative both at 
the inner and outer edges. Identical conclusions have been reached, but 
not explicitly stated, by Paczy{\'n}ski (1998): see his equation (20) 
which is equivalent to our equation (6.6).  
 
From equation (5.2) describing a typical angular momentum distribution in 
the disc, and equations (6.5), (6.6) one may calculate $B(r)$ in the 
whole disc. Figure~2 shows the Bernoulli function by the thick solid 
line. The thin broken line shows the prediction of the self-similar 
model. In the same Figure~2 we present for comparison 
by the thin solid line the time-averaged
Bernoulli function for 2-D numerical model of ADAF with $\alpha=0.01$.
Details of numerical technics described by Igumenshchev \& Abramowicz (1999).
The model has $r_{in}=3 r_G$, $r_{out}=8000 r_G$, and $\gamma=5/3$.

%%%%%%%%%%%%%%%%%%%%%%%%%%%%%%%%%%%%%%%%%%%%%%%%%%%%%%%%%%%%%%%%%%%%%%%%%%%% 
%%%%%%%%%%%%%%%%%%%%%%%%%%%%%%%%%%%%%%%%%%%%%%%%%%%%%%%%%%%%%%%%%%%%%%%%%%%% 
\begin{figure} 
%\vspace{8.2cm} 
\hbox{\psfig{file=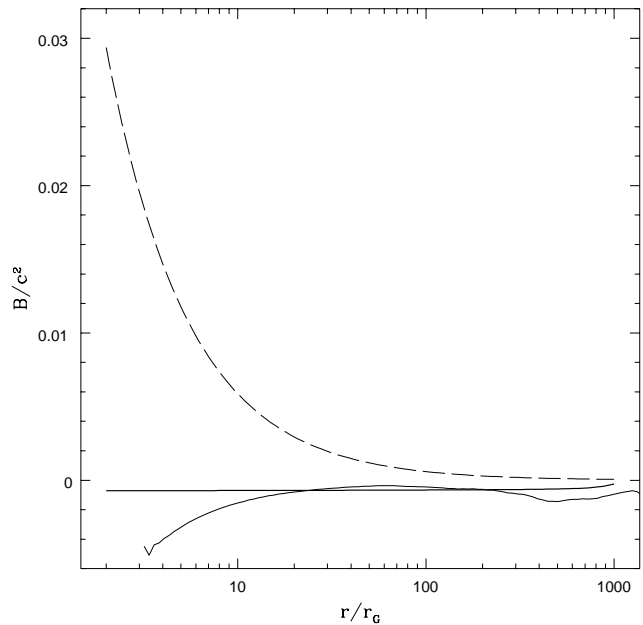,height=8.2cm}} 
\caption{ 
The Bernoulli function in ADAFs with small viscosity is 
everywhere negative. The solid thick line corresponds to the 
angular momentum distribution given in Figure~1, 
with $r_{in} = 2.003 r_G$, $r_{out} = 10^3 r_G$. 
The thin dashed line corresponds to the Narayan \& Yi (1994) 
self-similar solution (with $\gamma=3/2$), 
not compatible with the boundary conditions, used by BB99. 
The thin solid line repersents the time-averaged Bernoulli function
for 2-D numerical model of ADAF with $\alpha=0.01$.
} 
 \label{fig2} 
\end{figure} 
%%%%%%%%%%%%%%%%%%%%%%%%%%%%%%%%%%%%%%%%%%%%%%%%%%%%%%%%%%%%%%%%%%%%%%%%%%%% 
%%%%%%%%%%%%%%%%%%%%%%%%%%%%%%%%%%%%%%%%%%%%%%%%%%%%%%%%%%%%%%%%%%%%%%%%%%%% 
 
One concludes that in ADAFs with small viscosity, which fulfill  standard 
boundary conditions, the Bernoulli function must be everywhere negative. 
This conclusion is independent of the functional form of viscosity. 
Obviously, a flow which has the Bernoulli function that is everywhere 
negative does not experience outflows. 
Note, however that $B(r)$ calculated here is averaged with respect to
the flux of mass. Thus, it may happen that $B>0$ in some polar directions.
This is indeed the case close to the disc surface of some low viscosity 
numerical models calculated by ICA96. These models show weak outflows,
with $\dot{M}_{out}\ll\dot{M}_0$.

\section{Numerical simulations} 
 
Analytic arguments presented in the previous Section and pointing out 
that no significant outflows of hydrodynamical origin should be present 
in low viscosity ADAFs have been fully confirmed by all numerical 
simulations performed to date. Below we give a list of some  
representative works. 
 
\noindent 
{\it 1-D global simulations of transonic flows in optically thick case 
(slim discs)}: Abramowicz, Czerny, Lasota \& Szuszkiewicz (1988); Kato, 
Honma \& Matsumoto (1988ab); Chen \& Taam (1993); Szuszkiewicz \& Miller 
(1997). 
 
\noindent 
{\it 1-D global simulations of transonic flows in the optically thin case 
(ADAFs)}: Honma (1996); Chen, Abramowicz \& Lasota (1997);  
Narayan, Kato \& Honma (1997); Gammie \& Popham (1998);  
Nakamura, Kusunose, Matsumoto \& Kato (1998); 
Igumenshchev, Abramowicz \& Novikov (1998); Ogilvie (1999). 
 
\noindent {\it 2-D time dependent simulations of ADAFs}: Igumenshchev, 
Chen \& Abramowicz (1996); Igumenshchev \& Abramowicz (1999); 
Stone, Pringle \& Begelman (1999).
 
%\noindent The only numerical simulations which shows outflows for low 
%viscosity ADAFs, are 2-D simulations by Stone, Pringle \& 
%Begelman (1999). 
%These simulations, however, do not reproduce the well-understood 
%stabilization mechanism by the Roche lobe overflow. They disagree with 
%analytic results discussed in previous Sections of this paper. They also 
%disagree with all other numerical simulations listed above, and with  
%numerical 2-D simulations by R\'o\.zyczka, Bodenheimer \& Bell (1994),  
%which concern a different astrophysical context but are relevant here,  
%and with the 
%analytic 3-D model of Klu{\'z}niak \& Kita (1999). This, of course, does not
%mean that 
%Stone et al. (1999) are wrong but shows that their simulations are 
%fundamentally different from other attempts to model accretion flows. 
%Stone, Pringle \& Begelman  
%It is not yet clear whether the 
%problems are purely technical, e.g. due to an incorrect detail of the 
%numerical treatment of the inner boundary condition, or whether they are 
%more fundamental.  
%This problem will be solved by a direct comparison of  numerical simulations  
%(Abramowicz, Igumenshchev \& Stone, 1999; in preparation). 

\section{ADAFs with large viscosity} 
 
We have seen that there is a significant theoretical and numerical 
evidence which shows that purely hydrodynamical 
outflows in ADAFs with low viscosity are unlikely to occur. Physical reasons 
for the absence of outflows in this case seem to be well understood. They 
depend on some fundamental properties of the black hole gravity. 
 
The situation in the case of moderate and large viscosity is less clear.   
The analytic calculation of $B(r)$ presented in the Section 5 cannot be 
repeated in the case of large viscosity because in this calculation one 
assumes that $V_r \ll V_{\phi}$, while in flows with large viscosity, all 
velocity components are of the same order. This brings an additional 
unknown term to the energy equation (6.1c) so that the system has too 
many unknowns to be solved. 
Also, one can not further assume zero viscous torque acting at $r_{in}$,
if non-circular motions are significant.
Thus, equations (6.1) are insufficient to 
calculate $B(r)$ even in the self-similar case. In the large viscosity 
case one can argue neither that self-similarity implies $B(r)>0$, nor 
that the boundary conditions imply $B(r)<0$. 
 
Purely hydrodynamic outflows have been seen in numerical 
2-D simulations of high viscous accretion flows  
(Igumenshchev \& Abramowicz 1999), and in 2-D self-similar 
models (Xu \& Chen 1997, and references therein), but they are not a 
universal property of viscous flows. From the existing results it 
is obvious that the presence of outflows depends on the magnitude of 
viscosity parameter $\alpha$, adiabatic index $\gamma$, and, probably, on the outer boundary conditions at $r_{out}$. However, the 
boundary ${\cal F}(\alpha, \gamma, ...)$ that divide the 
parameter space into the flows with and without outflows is yet to be
found. 
 
Certainly, one cannot argue that $B > 0$ implies outflows. We illustrate 
this point by showing in Figure~3 a model of ADAF ($\alpha = 0.3$, 
$\gamma = 3/2$, $r_{in}=3 r_G$, $r_{out} = 8000r_G$) in which  $B>0$  
%(for $B$ been averaged over the polar angle $\theta$) 
and no outflows.    
The model has been calculated using numerical technics described 
by Igumenshchev \& Abramowicz (1999).
In Figure~4, for the same model, we show calculated distributions of 
$B(r)$ (solid line),
%%%%%%%%%%%%%%%%%%%%%%%%%%%%%%%%%%%%%%%%%%%%%%%%%%%%%%%%%%%%%%%%%%%%%%%%%%%%%%
$$
{B}(r)=
{2\pi r^2\over\dot{M}_0 c^2}\int_0^\pi\rho V_r\left({1\over 2}V^2+
  W-{GM\over r}\right)\cos\theta d\theta,
\eqno(8.1) $$
%%%%%%%%%%%%%%%%%%%%%%%%%%%%%%%%%%%%%%%%%%%%%%%%%%%%%%%%%%%%%%%%%%%%%%%%%%%%%%
normalized viscous energy flux (dashed line),
%%%%%%%%%%%%%%%%%%%%%%%%%%%%%%%%%%%%%%%%%%%%%%%%%%%%%%%%%%%%%%%%%%%%%%%%%%%%%%
$$
{G}(r)=-{2\pi r^2\over\dot{M}_0 c^2}\int_0^\pi\left(
  V_r\Pi_{rr}+V_\theta\Pi_{r \theta}
  +V_\phi\Pi_{r \phi}\right)\cos\theta d\theta,
\eqno(8.2) $$
%%%%%%%%%%%%%%%%%%%%%%%%%%%%%%%%%%%%%%%%%%%%%%%%%%%%%%%%%%%%%%%%%%%%%%%%%%%%%%
and normalized  total energy flux (solid thick line),
%%%%%%%%%%%%%%%%%%%%%%%%%%%%%%%%%%%%%%%%%%%%%%%%%%%%%%%%%%%%%%%%%%%%%%%%%%%%%%%
$$
{\dot{E}_0(r)\over\dot{M}_0 c^2}={B}(r)+{G}(r).
\eqno(8.3) $$
%%%%%%%%%%%%%%%%%%%%%%%%%%%%%%%%%%%%%%%%%%%%%%%%%%%%%%%%%%%%%%%%%%%%%%%%%%%%%%%
In (8.1)-(8.3) $r$, $\theta$ and $\phi$ are spherical coordinates and
${\bf\Pi}$ is the shear stress tensor.
In the model the inward energy advection flux
[term $B$ in (8.3) and short-dashed line in Fig.~4]
almost equals to the outward viscous energy flux
[term $G$ in (8.3) and long-dashed line in Fig.~4] at each radius.
This behaviour is similar to that in 
the self-similar Narayan \& Yi (1994) solution, where the
oppositely directed fluxes exactly compensate.
For comparison,
we show $B(r)$ for the self-similar solution in Figure~4 by dotted line.
When correct boundary conditions are taken into account
there is no exact compensation and 
the total energy flux $\dot{E}_0$ must be a small (in absolute value)
non-positive constant
in the stationary dissipative accretion flow of the type discussed here. 
Note that in the actual model $\dot{E}_0$ 
(solid line in Fig.~4) is not constant and oscillates
with a small amplitude
due to an inaccuracy of our numerical scheme,
which does not exactly conserve the total energy balance.
This inaccuracy is inside $\approx 5\%$ of relative error, which
is acceptable for our purposes.

%%%%%%%%%%%%%%%%%%%%%%%%%%%%%%%%%%%%%%%%%%%%%%%%%%%%%%%%%%%%%%%%%%%%%%%%%%%% 
%%%%%%%%%%%%%%%%%%%%%%%%%%%%%%%%%%%%%%%%%%%%%%%%%%%%%%%%%%%%%%%%%%%%%%%%%%%% 
\begin{figure} 
%\vspace{8.2cm} 
\hbox{\psfig{file=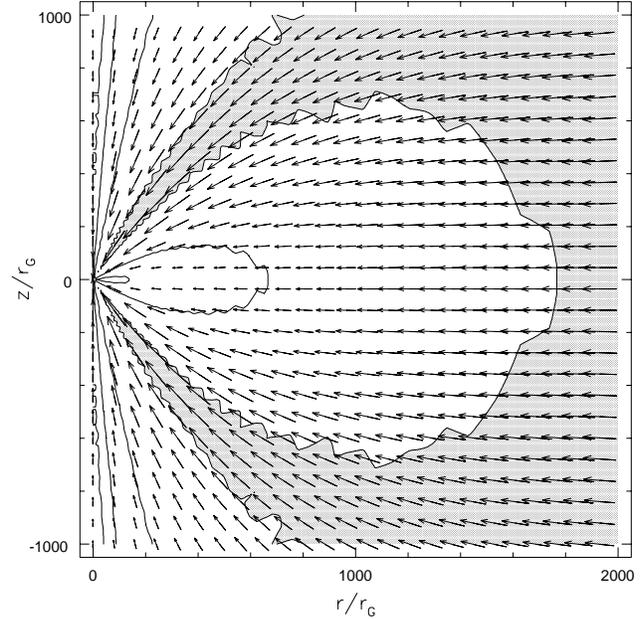,height=8.2cm}} 
\caption{2-D numerical stationary ADAF model with $\alpha=0.3$, $\gamma=3/2$, 
$r_{in}=3 r_G$ and $r_{out}=8000 r_G$. 
Only inner part of the model is shown. 
Arrows indicate the magnitude of the vector 
$r^2\rho\vec{v}$, where $r$ is the spherical radius. 
Regions with positive Bernoulli function are in white, while 
regions with negative values are in grey. 
Clearly, $B>0$ is not sufficient for outflows. 
} 
 \label{fig3} 
\end{figure} 
%%%%%%%%%%%%%%%%%%%%%%%%%%%%%%%%%%%%%%%%%%%%%%%%%%%%%%%%%%%%%%%%%%%%%%%%%%%% 
%%%%%%%%%%%%%%%%%%%%%%%%%%%%%%%%%%%%%%%%%%%%%%%%%%%%%%%%%%%%%%%%%%%%%%%%%%%% 

%%%%%%%%%%%%%%%%%%%%%%%%%%%%%%%%%%%%%%%%%%%%%%%%%%%%%%%%%%%%%%%%%%%%%%%%%%%%
%%%%%%%%%%%%%%%%%%%%%%%%%%%%%%%%%%%%%%%%%%%%%%%%%%%%%%%%%%%%%%%%%%%%%%%%%%%%
\begin{figure}
%\vspace{8.2cm}
\hbox{\psfig{file=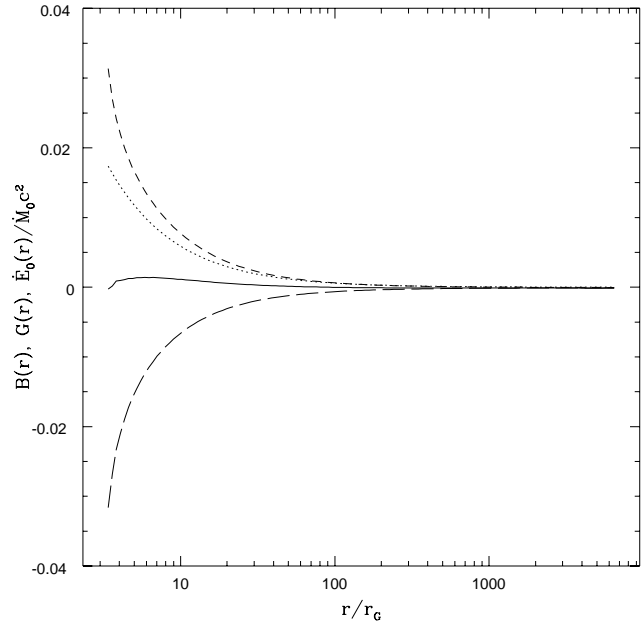,height=8.2cm}}
\caption{
Radial profiles of $\theta$-averaged quantities, given by eqs (8.1)-(8.3),
for the model presented in Figure~3.
Profiles of ${B}(r)$ (short-dashed line), 
${G}(r)$ (long-dashed line) and ${\dot{E}}_0(r)/\dot{M}_0 c^2$ 
(solid line)
are shown. The dotted line corresponds to $B(r)$ for  
the self-similar Narayan \& Yi (1994) solution.
See text for explanations.
}
 \label{fig4}
\end{figure}
%%%%%%%%%%%%%%%%%%%%%%%%%%%%%%%%%%%%%%%%%%%%%%%%%%%%%%%%%%%%%%%%%%%%%%%%%%%%
%%%%%%%%%%%%%%%%%%%%%%%%%%%%%%%%%%%%%%%%%%%%%%%%%%%%%%%%%%%%%%%%%%%%%%%%%%%% 

\section{Conclusions} 
 
\noindent 
1. Significant outflows 
of purely hydrodynamical origin are not present in ADAFs 
with low ($\alpha \la 0.1$) viscosity. This conclusion follow from 
general theoretical arguments that involve fundamental properties of 
black hole gravity, and are well understood. All numerical simulations 
%except one (Stone et al. 1999) 
confirm this theoretical 
arguments and show no outflows. 
 
%\noindent 
%2. One should understand the reason why the numerical simulations by 
%Stone et al. (1999) are the only ones showing outflows at low viscosity 
%(Abramowicz et al. 1999, in preparation). 
 
\noindent  
2. It would be very interesting to perform a systematic investigation in 
the parameter space \{$\alpha$, $\gamma$, $r_{\rm out}$\} 
by filling it with models of ADAFs and find the regions (necessarily with  
$\alpha > 0.1$) corresponding to outflows.  
%(Igumenshchev et al. 1999, in preparation). 
 
\noindent 
3. Outflows from ADAFs might occur due to non-hydrodynamical factors such as 
magnetic fields, radiation, etc. These processes cannot be modeled in 
purely hydrodynamical terms by adopting special value of $\gamma$ or form of 
the Bernoulli function: the extra physics should enter 
through solutions of the relevant equations (see e.g. King \& Begelman 1999). 
 
\subsection*{Acknowledgments} 
MAA and JPL are grateful to Mitch Begelman,  
Roger Blandford, Shoji Kato, Ramesh Narayan, Bohdan Paczy\'nski,  
Martin Rees and Ron Taam for stimulating discussions during  
the Santa Barbara ITP program `Black Hole in Astrophysics'.  
This research was supported in part by the 
National Science Foundation under Grant No. PHY94-07194.

\label{lastpage} 
%%%%%%%%%%%%%%%%%%%%%%%%%%%%%%%%%%%%%%%%%%%%%%%%%%%%%%%%%%%%%%%%%%%%%%%%%% 
%%%%%%%%%%%%%%%%%%%%%%%%%%%%%%%%%%%%%%%%%%%%%%%%%%%%%%%%%%%%%%%%%%%%%%%%%% 
 
\end{document}